\documentclass[12pt]{iopart}

\usepackage[dvips]{graphicx}
\usepackage{color}

\newcommand{\gtrsim}{\stackrel{>}{\sim}}

\begin{document}

{\color{red}Published in Nanotechnology {\bf 20}, 325602 (2009)}

\title[Atomic layer deposition of ZnS nanotubes]{Atomic layer deposition of ZnS nanotubes} 
\author{Sh. Farhangfar$^1$, R. B. Yang$^2$, M. Pelletier$^3$, K. Nielsch$^1$}  

\address{$^1$Institute of Applied Physics, University of Hamburg, Jungiusstr. 11, D-20355 Hamburg, Germany} 
\address{$^2$Max Planck Institute of Microstructure Physics, Weinberg 2, D-06120 Halle, Germany}
\address{$^3$Institute of Physical Chemistry, University of Hamburg, Bundesstr. 45,
D-20146 Hamburg, Germany}
\ead{sfarhang@physik.uni-hamburg.de}
\date{\today}
\begin{abstract}
We report on  growth of high-aspect-ratio ($\gtrsim300$)   zinc sulfide nanotubes with variable, precisely tunable, wall thicknesses and tube diameters into highly ordered pores of anodic alumina templates by atomic layer deposition (ALD) at temperatures as low as $75$ $^{\circ}$C.  Various characterization techniques are  employed to gain information on the composition, morphology, and crystal structure of the synthesized samples.  Besides practical applications, the ALD-grown tubes could be envisaged as model systems for the study of a certain class of size-dependent quantum and classical  phenomena.    
\end{abstract}

\submitto{\NT}

\maketitle


Functional nanostructures are currently subjects  of increasing interest. This is mainly due to the recent advances in the nanofabrication techniques and a better control on the confinement geometry and growth conditions of the low-dimensional systems. In this respect, semiconducting nanowires are of particular importance. On the one hand, they can be utilized in highly efficient  optical, electronic, biological, and sensing devices and, on the other hand, they can equally be exploited as model systems for the study of generic solid-state phenomena \cite{law04,lieber07}.

Here, we report on highly size-controlled synthesis of zinc sulfide (ZnS) nanotubes (NTs) by atomic layer deposition (ALD) and characterization of the tubes thus obtained. By selection of suitable precursors  and through utilization of surface chemistry reactions, ZnS nanotubes are grown  onto the pore walls of perfectly ordered porous anodic alumina templates. The diameter and length of  the hollow pores in the anodic alumina membrane which serves as a template for the tubes can be adjusted by optimization of  the anodization conditions in the process of template fabrication. The wall thickness of the tubes, subsequently, is defined through the chemical deposition parameters such as the type of the used precursors, the growth temperature, and the deposition time.

Bulk zinc sulfide is a group II-VI direct wide-bandgap semiconductor. At room temperature, it has a fundamental energy gap of about $3.7$ eV between ${\it\Gamma}_{15}$$-$${\it\Gamma}_{1}$ symmetry points in its cubic crystal (zincblende) phase. The corresponding value for the less frequently appearing hexagonal form (wurtzite) is $3.9$ eV \cite{landolt99}.  In contrast to most other binary semiconductors, ZnS can be doped both by donors and by acceptors to lead to either $n$- or $p$-type electrical conduction. Due to its optical properties, it is also a material of choice in certain optoelectronic applications. These include, among the others, exploitation of ZnS as a phosphor or scintillation material, or its use in fabrication of infrared filters. Still, contrary to its closest counterpart, ZnO, which has been studied extensively in the  nanoscales \cite{mende07}, investigations on  zero- and one-dimensional ZnS nanostructures   have been relatively scarce to date.  Recent progress in this direction involves fabrication of ZnS nanowires by  intermittent laser ablation-catalytic technique \cite{jiang03}, exploitation of single-source molecular precursors \cite{barrelet03}, pulsed laser vaporization \cite{xiong04}, vertical catalytic growth on alumina templates \cite{ding04}, as well as chemical vapor deposition   \cite{moon06}.  Correspondingly, ZnS tubular structures have been synthesized by utilization of  ZnO nanowires and their chemical manipulation \cite{xwang02}, thermo-chemical process  \cite{zhu03}, chemical growth of ZnS films onto carbon nanotube templates \cite{hzhang04}, surfactant micelle-template induced reactions  \cite{lv04}, solid-state chemical reactions \cite{zhou04},  high temperature chemical vapor deposition (CVD)  \cite{shen05},  metal organic chemical vapor deposition  \cite{zhai06}, and by hydrothermal methods \cite{shi07}. Furthermore, alternative avenues, such as rolling of the certain layered materials to the tubular forms, have been explored recently \cite{yanetal08} and can possibly be examined for the fabrication of ZnS nanotubes as well.  Yet,  most of the above-mentioned approaches result in growth of bundles, may involve high temperature reactions, or have their strong limitations with respect to the accurate size and distribution control of the synthesized  structures. Also, the majority of these techniques can only be utilized to synthesize either wire or tube geometries, but not the both.  Here, we report for the first time on  fabrication of ZnS nanotubes by ALD. This approach can be equally used to obtain nanowires. In what follows, we will first explain ZnS nanotube fabrication process  and will then introduce the results of the experimental characterizations performed on the obtained structures.    

First, to optimize deposition conditions in the ALD process, we examined SiO$_x$/Si wafers and commercially available anodic aluminum oxide (AAO) membranes. The latter had a distribution of pore diameters in the range 30$-$300 nm. The experiments were further carried out by use of home-made highly ordered alumina matrices, as is  discussed below.

AAO templates were electrochemically fabricated by anodizing pure Al in acidic electrolytes. Highly ordered pores with average diameters of about $30$, $45$, and $200$ nanometers were obtained by use of sulfuric, oxalic, and phosphoric acid solutions, respectively. Precise control on the pore size and the interpore distances can be achieved by variation of electrolyte concentration, temperature, and the current density through anodized aluminum.  The length of pore channels, up to $100$ $\mu$m, was tailored by anodization time. More details on AAO template fabrication are given elsewhere \cite{masuda95,nielsch02}.

ALD is a bottom-up fabrication technique suitable for deposition of highly conformal thin films \cite{puurunen05,knez07}. In each self-limiting cycle of an ALD process, reactants in vapor phases are guided alternately into a vacuum chamber and are let to react on the exposed surface. Since, ideally, within each deposition cycle, only one monolayer is formed, the deposition will be highly conformal and its thickness can be accurately tailored  by defining the number of cycles. In the case of a nanotube geometry, this will lead to a homogeneous radial coating of the walls and will allow precise control over the formed NT wall thickness.  As precursors, we used diethylzinc, Zn(C$_{2}$H$_{5}$)$_{2}$, (purchased from Strem) and hydrogen sulfide, H$_2$S, (from Sigma Aldrich). Diethylzinc was chosen over the other widely used precursors, such as zinc chloride [ZnCl$_2$], zinc acetate [Zn(CH$_3$COO)$_2$], or zinc oxy-acetate [Zn$_4$O(CH$_3$COO)$_6$], for two reasons. On the one hand, its use is less prone to lead to the aging phenomena encountered in some practical applications \cite{stuyven02}, on the other hand it has smaller molecules than those of most other metalorganics.  The latter implies a better diffusivity along the narrow pore channels in AAO membranes and facilitates fabrication of high-aspect-ratio structures.   Diethylzinc  was kept in a stainless steel bottle with an air-tight valve at room temperature. Hydrogen sulfide  was supplied by a pressurized canister regulated at 0.3$-$0.7 bar. The ALD process consisted of alternating and separate introductions of Zn(C$_2$H$_5$)$_2$ and H$_2$S into the reaction chamber.   
For each precursor, the pump was first disconnected from the chamber and the vapor was  allowed to enter into it ("pulse"). The precursor vapor was then left to react in the chamber with the AAO substrate ("exposure"), after which the chamber was evacuated ("purge"). The three steps were carried out for the second precursor, and the whole process was repeated in a cyclic manner (see Fig. 1).   The precursors were kept at room temperature. The pulse, exposure, and purging durations were, respectively, $50$ ms, $20$ s, and $40$ s for both  the precursors. Within each cycle, the reactant gases were taken into and out of the chamber by a $40$ cm$^3$/min  flow of argon carrier gas. During the purging steps, the chamber was pumped down to a vacuum pressure better than $0.5$ mbar. The deposition temperature (temperature of the AAO template) was set by regulating the heating power through the hot plate of the ALD chamber.  Figure 2 shows variation of the radial growth rate of the nanotubes with the temperature. Here, first a series of $300$ ALD cycles at 
$T=75, 100, 120, 150$, and 180 $^\circ$C were performed to obtain tubes with wall thicknesses of about $40$ nm or larger. The precise thickness of each set of depositions was then estimated both by a field-emission scanning electron microscope (FE-SEM) and by a transmission electron microscope (TEM). Obtained nanotubes with lengths as long as 50 $\mu$m did not exhibit any traceable variation in their wall thickness distribution along the tube axes. This fact is an indication of the good diffusivity of the precursor molecules along the long pore channels of AAO templates and points at feasibility of fabrication of nanotubes with even larger aspect ratios than those achieved in the present study, $\sim300$. Still, while the depositions at temperatures up to 150 $^\circ$C resulted in very smooth and conformal film surfaces, the one at 180 $^\circ$C  gave rise to the formation of rough and somewhat granular films. Figures 3 and 4, taken by a FE-SEM, exhibit examples of the nanotubes formed by atomic layer deposition of ZnS thin films  onto the pore walls of host templates with various pore sizes  at different temperatures. The relatively large  growth rates here, about 1.4$-$2.0 {\AA}/cycle, and the deposition temperatures as low 75 $^\circ$C, have to be compared to those reported for the ALD  of ZnS thin films onto glass substrates earlier \cite{stuyven02} (less than 0.9  {\AA}/cycle at 200$-$300 $^\circ$C, and a very strong dependence of the growth rate on $T$), and to the lattice constants of crystalline ZnS in its zincblende and wurtzite phases, 5.41 {\AA} and 3.82 {\AA}, respectively \cite{landolt99}.

To facilitate further studies on the physical properties of the synthesized tubes, it is of crucial importance to be able to detach them from the template host. To get a solution of single ZnS nanotubes, a small piece, $\sim10$ mm$^2$ in size, of embedding AAO matrix was dissolved in 1 ml of a one-molar NaOH solution and was subsequently purified to neutrality in distilled water. (The pore density of the AAO templates is in the range ${10^6}-{10^8}$ mm$^{-2}$.)  It is worth mentioning that the density of nanotubes in such a suspension would be large enough to allow for, for instance, optical measurements on the samples \cite{optical}.  Figure 5 shows an image of the tubes dispersed on a SiO$_x$/Si substrate taken by a FE-SEM. The inset is a transmission electron microscope  micrograph taken from a mixture of tubes with different diameters.

Elemental composition of the samples was determined by an energy dispersive X-ray (EDX) spectrometer integrated into a SEM. The ratio of measured stoichiometric quantities, $30.93\%$ zinc and $33.81\%$ sulfur, is close to the expected 1$:$1 value (see Fig. 6). Visible traces of  Si and O originate from the substrate wafer and those of Al and O are due to the remnants of the dissolved alumina template. The detected large amount of carbon, instead, has a less certain origin. To clarify whether it is an  artifact of the EDX setup (typically encountered with in the detection of light elements) or it originates from the incomplete reaction of the precursors, the samples were further analysed by alternative complementary techniques \cite{kolbe}. These included CHNOS analysis (burning of the samples and analysing of the product gases) for the determination of carbon, sulfur, and oxygen contents, atomic absorption spectroscoy (AAS) for the detection of zinc, inductively coupled plasma (ICP) technique for the verification of phosphorus and aluminum, as well as photometry for the  determination of silicon content. Table 1 summarizes the outcomes of these analyses. Evidently, the stoichiometric ratios of zinc and sulfide are correct and there is only a negligible trace of carbon or other impurities present. Interestingly, and somewhat counter-intuitively, the film deposited at 120 $^\circ$C has better quality (in terms of both the stoichiometry and the lesser amount of impurity carbon)  than that grown at 180 $^\circ$C.

Figure 7 shows a high-resolution (HR) TEM pattern taken from the wall surface of a ZnS nanotube. Clearly, a larger portion of the surface is amorphous. Yet, some crystalline domains are distinguishable.  
A TEM diffraction pattern taken from such a domain is presented in Fig. 8. This pattern corresponds to that of polycrystalline ZnS in its cubic phase. Nevertheless, as  the  distribution of such domains along the NT wall surface is irregular,  one cannot unequivocally define the growth directions of the films. Indeed, poor crystalline quality of the ALD coatings is a common feature in most ALD conditions and can hardly be overcome during the growth procedure itself. However, a post-annealing of such films in an inert atmosphere may lead to certain improvements in some cases, as has recently been verified  for the case of zinc sulfide NTs synthesized by the high temperature CVD technique \cite{shen05}.

To summarize, we have fabricated  ZnS nanotubes  by low-temperature atomic layer deposition (ALD). The tubes had a very smooth wall surface and their diameter (30$-$200 nm), wall thickness (10$-$60 nm), and  their length (up to several tens of $\mu$m), were precisely tailored by a combination of electrochemical and ALD techniques. In addition, the effect of the deposition temperature on the growth rate of the films, chemical composition of the samples, as well as their crystal structure, were addressed.

We would like to thank Christian Klinke (Institute of Physical Chemistry, Hamburg) for the EDX data and Klaus-Peter Meyer (Max Planck Institute, Halle) for the RCA cleaning of the silicon substrates. Financial support from  the German Ministry for Education and Research (BMBF, projects 03N8701 and 03X5519) is gratefully acknowledged.


\clearpage

\clearpage 
\begin{figure}\label{Fig.1}
\begin{center}
\end{center}
\begin{center}
\includegraphics[width=125mm] {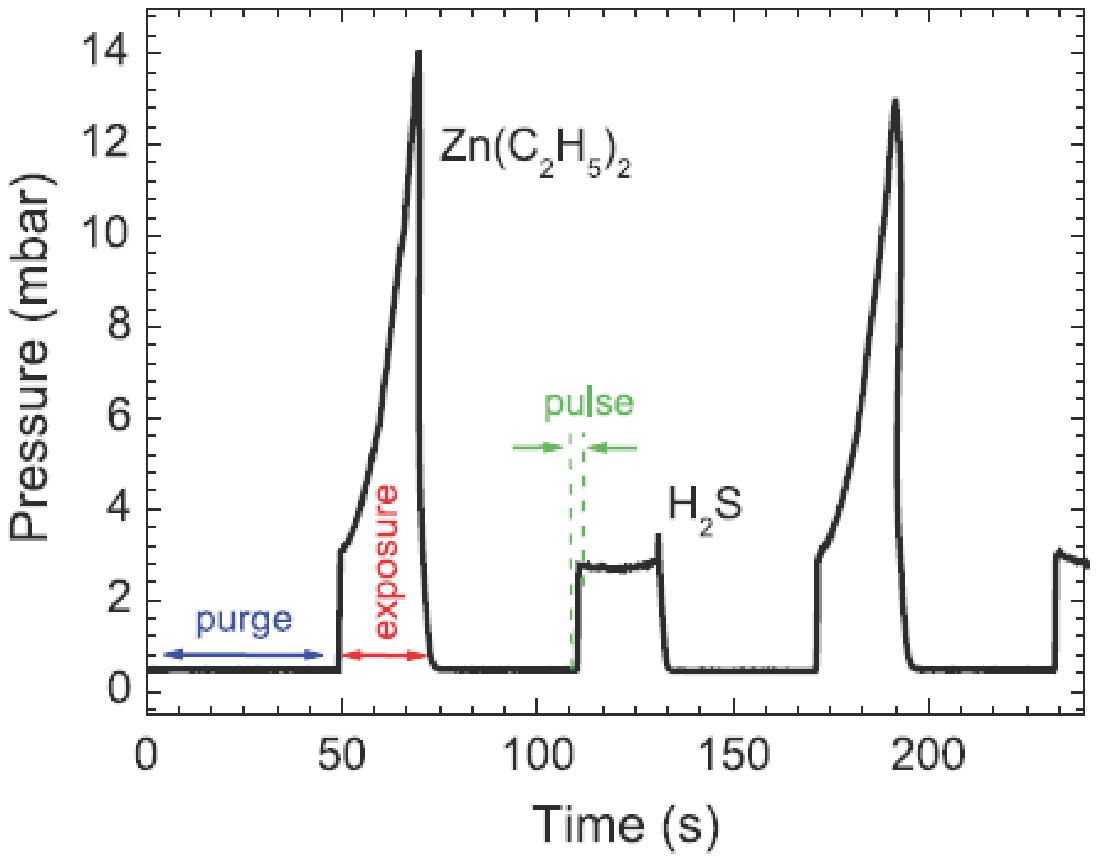}
\caption{A snapshot of the ALD cycles. The pulse, exposure, and purge periods for both the precursors diethylzinc and hydrogen sulfide were 50 ms, 20 s, and 40 s, respectively. Within the each purging step, the reaction chamber was pumped down to a pressure of about 0.4 mbar.}
\end{center}
\end{figure} 
\clearpage 
\begin{figure}\label{Fig.2}
\begin{center}
\end{center}
\begin{center}
\includegraphics[width=125mm] {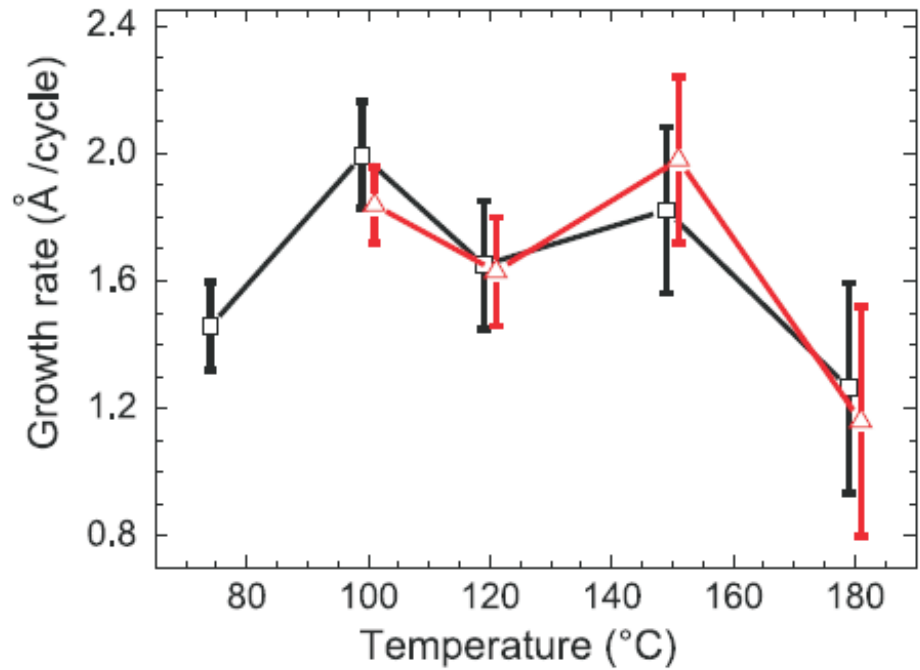}
\caption{Growth rates of the ALD process determined by a FE-SEM (open squares) and by  a TEM (open triangles) at $T=75, 100, 120, 150$, and 180 $^{\circ}$C. The solid lines are guides to the eye. Note that as the temperature rises, the films become rougher.}  
\end{center}
\end{figure} 
\clearpage 
\begin{figure}\label{Fig.3}
\begin{center}
\includegraphics[width=100mm] {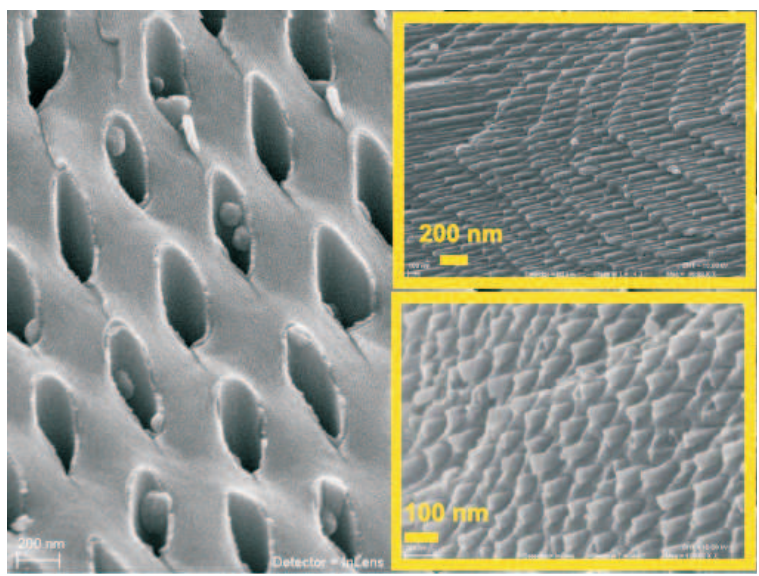}
\caption{FE-SEM micrographs of the ZnS nanotubes grown by ALD into ordered pores of an alumina template. The deposition was performed at 120 $^\circ$C.  The pores are $200$ nm wide. The insets show  narrower (30$-$45 nm) NTs.  To enhance the contrast, the outer surface of the membranes were wet-etched in a $5\%$ phosphoric acid solution.}  
\end{center}
\end{figure} 
\clearpage 
\begin{figure}\label{Fig.4}
\begin{center}				
\includegraphics[width=100mm] {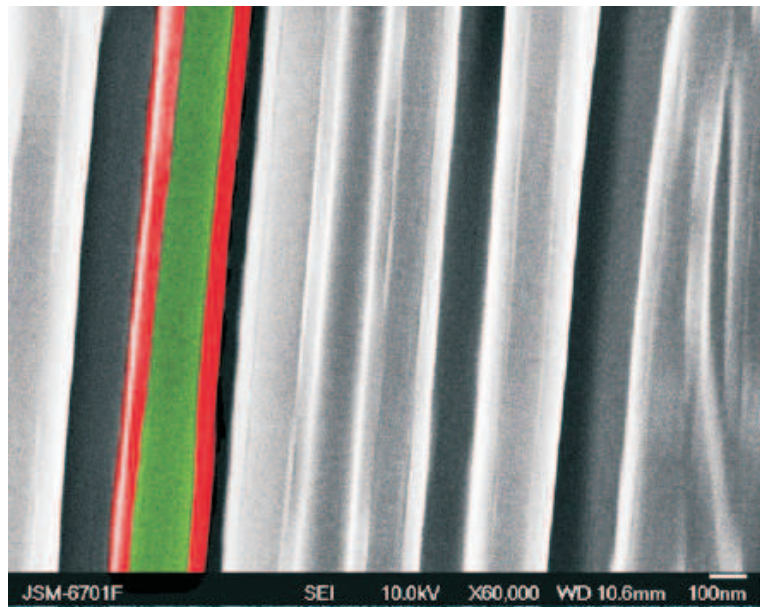}
\caption{Cross-section of the pore channels (green or gray) in an AAO template coated with uniform films of ZnS (red or light gray). The deposition was carried out at 75 $^{\circ}$C.}  
\end{center}
\end{figure}
\clearpage 
\begin{figure}\label{Fig.5}
\begin{center}				
\includegraphics[width=100mm] {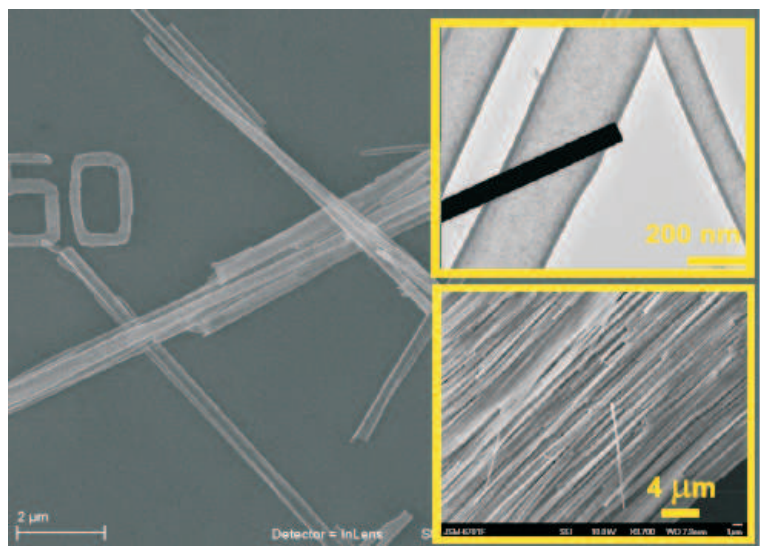}
\caption{Dispersed ZnS nanotubes on a SiO$_x$/Si substrate. The numbers are lithographically fabricated marks for addressing of individual tubes. The upper inset is a TEM image taken from a mixture of NTs with different diameters. The deposited wall thickness, $\sim17$ nm, is a result of 100 ALD cycles at 120 $^{\circ}$C. The lower inset depicts a bundle of ZnS nanotubes with lengths in excess of 50 $\mu$m deposited at 75 $^\circ$C.}  
\end{center}
\end{figure}
\clearpage 
\begin{figure}\label{Fig.6}
\begin{center}
\includegraphics[width=125mm] {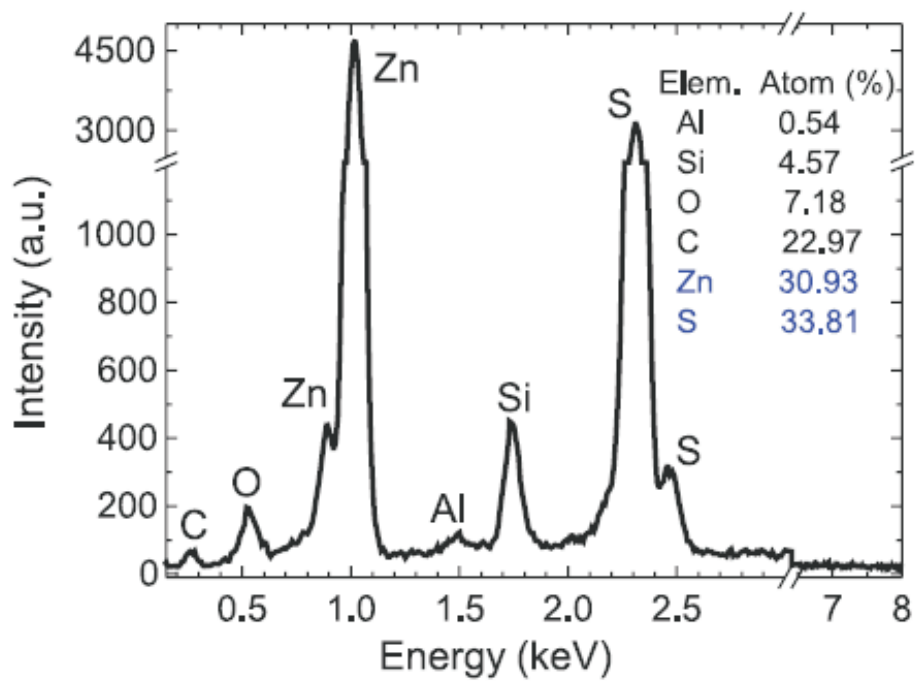}
\caption{EDX data of the nanotubes dispersed on a SiO$_x$/Si substrate. The detected elements and their respective portions are given on the graph. The high mass percentage of carbon is due to the typical uncertainties of the EDX technique in detection of light elements (see, Table 1). The ratio Zn:S is in good agreement with the ideal 1:1 stoichiometric value.}  
\end{center}
\end{figure}
\clearpage 
\begin{table}\label{Tabel.1}
\caption{\label{arttype}Elemental analysis of the films deposited onto Si substrates. Shortly prior to the deposition, the substrate surface was cleaned in a full RCA (Radio Corporation America) process. Note that the detected mass percentages of zinc and sulfur further confirm the desired 1:1 stoichiometric value for ZnS (cf. Fig. 6). For details, see the text.}

\begin{tabular*}{\textwidth}{@{}l*{15}{@{\extracolsep{0pt plus12pt}}l}}
\br
Element&$T=120$ $^{\circ}$C&$T=180$ $^{\circ}$C&Used technique\\
\mr
\verb"C"&$<6$ ppm&48 ppm&CHNOS (burning)\\
\verb"Zn"&1203 ppm (0.12\%)&1481 ppm (0.14\%)&AAS\\
\verb"S"&587 ppm (0.06\%)&585 ppm (0.06\%)&CHNOS\\
\verb"P"&$<1$ ppm&$<1$ ppm&ICP\\
\verb"Si"&99.61\%&99.47\%&Photometry\\
\verb"Al"&$<1$ ppm&$<1$ ppm&ICP\\
\verb"O"&912 ppm (0.09\%)&1365 ppm (0.14\%)&CHNOS\\
\br
\end{tabular*}
\end{table}

\clearpage
\begin{figure}\label{Fig.7}
\vspace{2cm}
\begin{center}
\includegraphics[width=125mm] {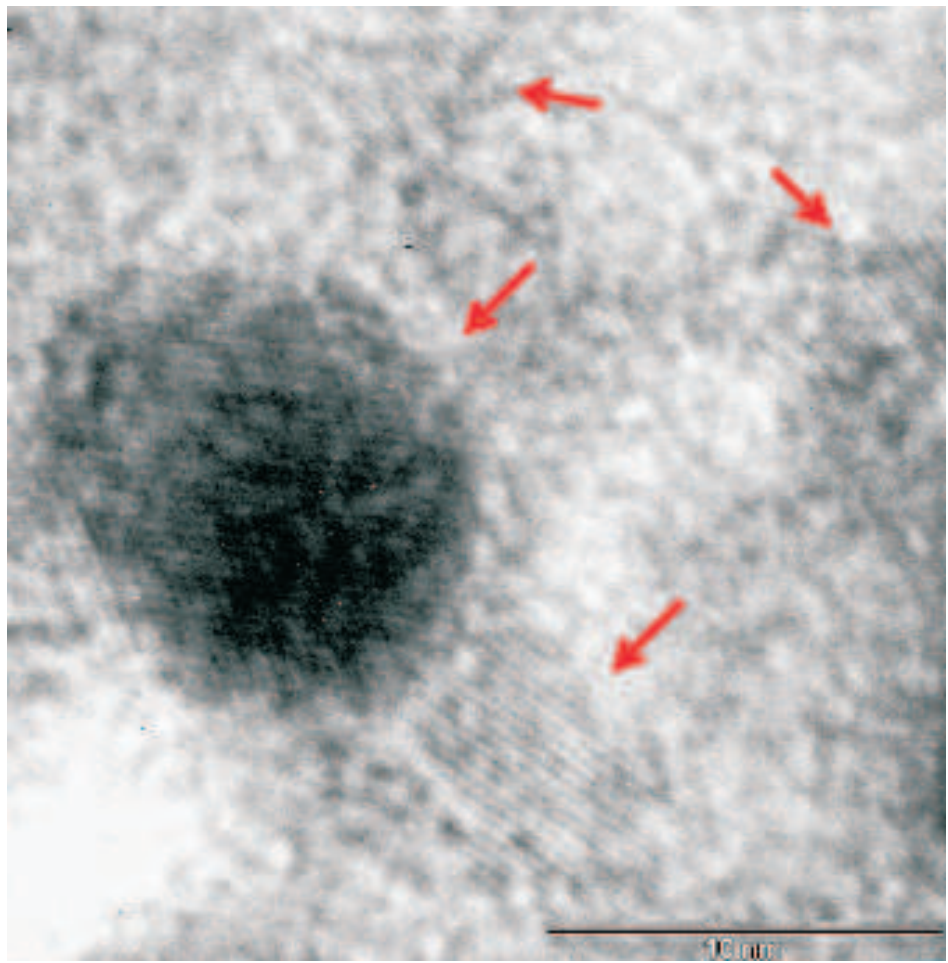}
\caption{HR-TEM image taken from the wall surface of a ZnS nanotube. Some crystalline domains are indicated by the arrows. The dark area on the left side was under exposure of an electron beam for a longer period of time and it corresponds to the TEM diffraction pattern presented in Fig. 8.}
\end{center}
\end{figure} 

\clearpage 
\begin{figure}\label{Fig.8}
\begin{center}
\includegraphics[width=100mm] {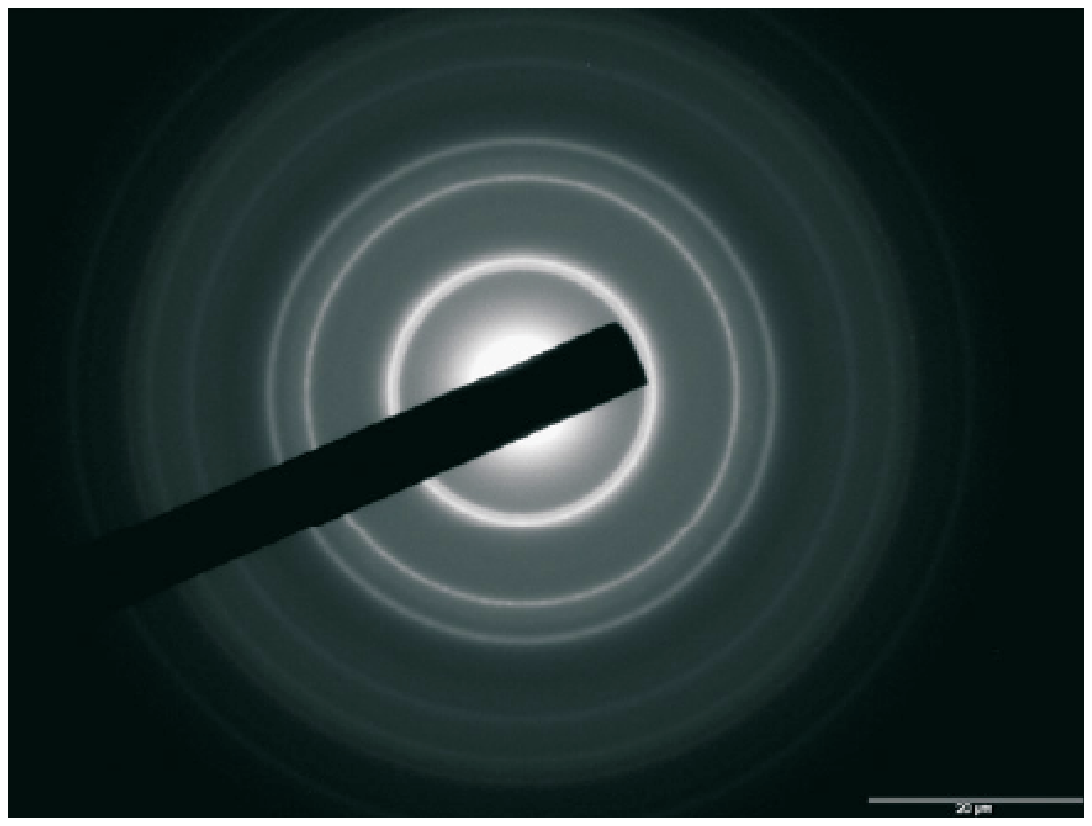}
\caption{A TEM diffraction pattern taken from a crystalline domain (indicated dark area in Fig. 7)  on the  nanotube wall. This pattern corresponds to that of a cubic  phase. The ring indices  are 111 (the innermost), 220 (the middle), and 311 (the outermost), respectively. Due to the inadequate resolution, the 200 ring is not visible.}  
\end{center}
\end{figure}


\section*{References}
\begin{thebibliography}{10} 

  

\bibitem{law04}  Law  Matt,  Goldberger  Joshua, and  Yang Peidong 2004 {\it Annu. Rev. Mater. Res.}  {\bf  122}  34. 

\bibitem{lieber07}  Lieber C and  Wang Z 2007 {\it MRS Bull.}  {\bf 32}  99. 




 

\bibitem{landolt99}  Landolt  H,  B\"ornstein R 1999 {\it Numerical data and functional relationships in science and technology, New Series III/23a} (Berlin: Springer-Verlag). 
  
\bibitem{mende07} For a recent review see, {\it e.g.},   Schmidt-Mende Lukas  and  MacManus-Driscoll  Judith L.  2007 {\it Mater. Today}  {\bf 10} 40. 


\bibitem{jiang03}  Jiang  Jang,  Meng  Xiang-Min,  Liu  Ji,  Hong  Zhi-Ro,  Lee  Chun-Sing, and  Lee Shuit-Tong  2003 {\it Adv. Mater.}  {\bf 15}  1195. 

\bibitem{barrelet03}  Barrelet  Carl J,  Wu  Yue,  Bell  David C, and  Lieber Charles  2003 {\it J. Am. Chem. Soc.}  {\bf 125}  11498. 


\bibitem{xiong04}  Xiong  Qihua,  Chen  G,  Acord  J  D,  Liu  X,  Zengel  J  J,  Gutierrez  H  R,  Redwing  J  M,  Lew Yan Voon  L  C,  Lassen  B, and  Eklund P  C  2004 {\it Nano Lett.}  {\bf 4} 1663. 

\bibitem{ding04}  Ding  J  X,  Zapien  J  A,  Chen  W  W,  Lifshitz  Y, and  Lee S  T  2004 {\it Appl. Phys. Lett.}   {\bf 85} 2361.



\bibitem{moon06}  Moon  Heesung,  Nam  Changhun,  Kim  Changwook,  Kim  Bongsoo 2006 {\it MRS Bull.}  {\bf 41} 2013.



\bibitem{xwang02}  Wang  Xudong,  Gao  Puxian,  Li  Jing,  Summers  Christopher J, and  Wang Zhong Lin 2002 {\it Adv. Mater.}  {\bf 14}  1732. 

\bibitem{zhu03}  Zhu  Ying-Chun,  Bando Yoshio  and  Uemera Yoichiro  2003 {\it Chem. Commun.}  836. 

\bibitem{hzhang04}  Zhang  Hua,  Zhang  Shuyuan,  Pan  Shuan,  Li Gongpu  and  Hou Jianguo  2004 {\it Nanotechnology}  
{\bf 15}  945. 

\bibitem{lv04}  Lv  R  T,  Cao  C  B,  Guo  Y  J,  Zhu H  S  2004 {\it J. Mater. Sci.}  {\bf 39}  1575. 

\bibitem{zhou04}  Zhou  Tao-Yu and  Xin  Xin-Quan 2004 {\it Nanotechnology}  {\bf 15}  534. 
 

\bibitem{shen05}  Shen  Xiao-Ping,  Han  Min,  Hong  Jian-Ming,  Xue  Ziling, and  Xu  Zheng   
2004 {\it Chem. Vap. Deposition}  {\bf  11}  250. 


\bibitem{zhai06}  Zhai  Tianyou,  Gu  Zhanjun,  Ma Ying,  Yang  Wansheng,  Zhao Liyum and  Yao  Jiannian   2006 {\it Mat. Chem. Phys.}  {\bf 100}  281.



\bibitem{shi07}  Shi  L,  Xu  Y   M,  Li  Quan,  Wu  Z  Y,  Chen  F  R, and  Kai  J  J   2007 {\it Appl. Phys. Lett.}  
{\bf 90}  211910.  

\bibitem{yanetal08} Yan Chenglin, Liu Jun, Liu Fei, Wu Junshu, Gao Kun, Xue Dongfeng 2008 {\it Nanoscale Res. Lett.} {\bf 3} 473.  

\bibitem{masuda95} Masuda H and Fukuda K 1995 {\it Science}  {\bf 268}  1466.

\bibitem{nielsch02} Nielsch Kornelius,  Choi Jinsub,  Schwirn  Kathrin,  Wehrspohn  Ralf B, and  G\"osele Ulrich  2002 {\it Nano Lett.}  {\bf 2}  677. 

\bibitem{puurunen05}  Puurunen Riikka L  2005 {\it J. Appl. Phys.}  {\bf 97}  121301. 

\bibitem{knez07}  Knez Mato,  Nielsch  Kornelius, and Niinist{\"o} Lauri   2007 {\it Adv. Matt.}  {\bf 19}  3425.

\bibitem{stuyven02} Stuyven Gert,  De Visschere Patrick,  Hikavyy  Andriy,  Neyts Kristiaan   2002 {\it J. Cryst. Growth}  {\bf 234}  690.  

\bibitem{optical} We are currently working on the optical characterization of the nanotube solutions. Based on our preliminary results, the tubes are semiconducting and their energy bandgaps are comparable to those of ZnS in its bulk phases. 

\bibitem{kolbe} The CHNOS, AAS, ICP, and photometry analyses were performed by the company  
{\it Mikroanalytisches Laboratorium Kolbe}: http://www.mikro-lab.de.  

\end{thebibliography}
\end{document}